\documentstyle[preprint,prb,aps]{revtex}
\draft

\def\vec#1{\ifcat #1 a \mathbf{#1} \else \boldsymbol{#1} \fi }

\begin{document}

\title{
\vspace*{5mm}
Marginal Fermi Liquid Theory in the Hubbard Model
\vspace{10mm}
}

\author{Y. Kakehashi and P. Fulde\\ \vspace{.5cm}}

\address{Max-Planck-Institut f\"ur Physik komplexer Systeme,
N\"othnitzer Str. 38, \\
D-01187 Dresden, Germany
\vspace{7mm}
} 

\date{\today  \vspace{3mm}}
\maketitle
\begin{abstract}
We find Marginal Fermi Liquid (MFL) like behavior in the
Hubbard model on a square lattice for a range of hole doping and on-site
interaction parameter $U$. Thereby we use a self-consistent projection
operator method. It enables us to compute the momentum and frequency dependence
of the single-particle excitations with high resolution. The Fermi surface is
found to be hole-like in the underdoped and electron-like in the overdoped
regime. When a comparison is possible we find consistency with finite
temperature quantum Monte Carlo results. We also find a discontinuous change
with doping concentration from a MFL to Fermi liquid behavior resulting from a
collapse of the lower Hubbard band. This renders Luttinger's theorem
inapplicable in the underdoped regime.
\vspace{10mm}
\end{abstract}


\nopagebreak

After the discovery of high-temperature superconductivity in layered Cu based
perovskites it was found that those materials exhibit quite unusual properties
in the normal state. For example, in underdoped materials, i.e., for hole
concentrations less than the one leading to the highest superconducting
transition temperature, the temperature dependent resistivity is found to be
$\rho(T)\sim T$ in the normal state. Also the nuclear relaxation rate, e.g., of
YBa$_2$Cu$_3$O$_7$ has an unusual temperature independent contribution. 

Aiming for an explanation of these strong deviations from a normal metal
behavior Varma et al. \cite{varma89} developed the Marginal Fermi Liquid
(MFL) theory. This theory assumes that the frequency $\omega$ and temperature
$T$ dependent self-energy $\Sigma(\omega, T)$ of the electrons behaves for
$\omega > T$ like Re$\Sigma(\omega, T) \sim \omega$ln$|\omega|$ and Im$\Sigma(\omega, T) \sim |\omega|$ in contrast to ordinary Fermi liquid theory
where Re$\Sigma(\omega, T) \sim \omega$ and Im$\Sigma(\omega, T) \sim
\omega^2$ holds. Note that at zero 
temperature the MFL form of the self-energy implies a diverging effective mass
at the Fermi energy. With these assumptions most of the observed strong
deviations from normal metal behavior could be explained surprisingly
well. However, the microscopic origin of MFL behavior of the self-energy has
remained an open problem.

There have been detailed studies of the two-dimensional (2D) Hubbard model as
a simple model for the high-T$_c$ cuprates \cite{dagotto94} mainly by using
advanced numerical techniques. We mention in particular the Lanczos method
\cite{dagotto94}, the Quantum Monte-Carlo (QMC)
\cite{bulut94,preuss94,grober00} method or calculations based on the Dynamical
Cluster Approximation (DCA) \cite{maier02}.

A perturbation analysis of the half-filled case at $T=0$ has shown that in the
weak Coulomb interaction limit a MFL type of self-energy is obtained
\cite{viro90}. It is due to the van Hove singularities which one is dealing
with in this particular case. But we know that electron correlations are strong
in the superconducting cuprates and that at $T=0$ the system is an
antiferromagnet \cite{dagotto94,grober00,jarrell01}. It is also known that by
hole doping the antiferromagnetic correlations are rapidly suppressed
\cite{dagotto94,grober00,maier02,onoda01}. Nevertheless MFL behavior continues
to exist in the underdoped regime and the question is whether or not it can be
explained within the 2D Hubbard model with fairly strong interactions.

The aim of this letter is to demonstrate that MFL behavior can indeed be
derived from a doped 2D Hubbard model on a square lattice at $T=0$ and large
on-site interaction. This has become possible with the help of a recently
developed self-consistent projection operator method (SCPM) \cite{kake04-1}. It
allows for high resolution calculations of the self-energy as regards its
momentum and energy and avoids certain problems previous numerical calculations
have had to face.

The SCPM is an extension to the nonlocal case of a projection operator coherent
potential approximation \cite{kake04-2}. The latter was shown to be
equivalent to the many-body CPA, the dynamical CPA as well as the dynamical
mean-field theory \cite{kake04-3}. In the following we outline briefly the main
equations which are used before we present the numerical results demonstrating
MFL behavior. More detailed derivations of the equations are found in the
original literature \cite{kake04-1,kake04-2,kake04-3}.

Starting point is the retarded Green function

\begin{equation}
G_{\mbox{\boldmath$k$}}(z) = \frac{1}{z - \epsilon_{\mbox{\boldmath$k$}} - \Lambda_{\mbox{\boldmath$k$}}(z)} \ .
\label{gk}
\end{equation}

\noindent Here $z=\omega + i \delta$ where $\delta$ is a positive infinitesimal
number, $\epsilon_{\mbox{\boldmath$k$}}$ is the Hartree-Fock one-electron dispersion
measured from the Fermi energy and $\Lambda_{\mbox{\boldmath$k$}}(z)$ 
is the self-energy
calculated from the nonlocal memory matrix $M_{ij}$ according to
 
\begin{equation}
\Lambda_{\mbox{\boldmath$k$}}(z)=U^{2}\sum_j M_{j0}(z) \exp (i \mbox{\boldmath$k$} \cdot \mbox{\boldmath$R$}_j) \ .
\label{lk}
\end{equation}

\noindent While $U$ denotes the Hubbard on-site interaction, 
$\mbox{\boldmath$R$}_j$ is
the position vector of site $j$. We calculate $M_{ij}(z)$ by using an
incremental cluster expansion up to two-sites contributions

\begin{equation}
M_{ii}(z) =  M^{(i)}_{ii}(z) + \sum_{l \neq i} \left(
M^{(il)}_{ii}(z)-M^{(i)}_{ii}(z)\right) \ , 
\label{icrii}
\end{equation}

\begin{equation}
M_{i \neq j}(z) = M^{(ij)}_{i \neq j}(z) \ , 
\label{icrij}
\end{equation}

\noindent where $M^{(i)}_{ii}(z)$ and $M^{(ij)}_{i \neq j}(z)$ are matrix
elements of the cluster memory matrices $M^{(c)}_{lm}(z)$ $(c=i, ij)$. The
latter can be expressed in terms of a ''screened memory matrix''
$\hat{\mbox{\boldmath$M$}}^{(c)}(z)$ and a matrix 
${\mbox{\boldmath$L$}}^{(c)}(z)$ which describes
on-site excitations. It is 

\begin{equation}
M^{(c)}_{lm}(z) = \left[ \hat{\mbox{\boldmath$M$}}^{(c)} 
\left( 1 - {\mbox{\boldmath$L$}}^{(c)} \cdot
\hat{\mbox{\boldmath$M$}}^{(c)} \right)^{-1} \right]_{lm} \ .
\label{memcij}
\end{equation}

The matrices have dimensions 1$\times$1 when $c=i$ and 2$\times$2 when
$c=(ij)$. Specifically $L^{(i)}(z)=U(1-2\langle n_{i-\sigma}\rangle)/[\langle
n_{i-\sigma} \rangle (1 - \langle n_{i-\sigma}\rangle)]$ while 
$\mbox{\boldmath$L$}^{(ij)}(z)$ is a diagonal matrix with elements 
$L^{(i)}(z)$ and
$L^{(j)}(z)$. As usual $\langle n_{i \sigma} \rangle$ is the average electron
number at site $i$ with spin $\sigma$. The screened memory matrix is calculated
from renormalized perturbation theory \cite{Lambda} as

\begin{equation}
\hat{M}^{(c)}_{ij}(z) = A_{ij} \int \frac{d\epsilon d\epsilon^{\prime} 
d\epsilon^{\prime\prime} \tilde{\rho}^{(c)}_{ij}(\epsilon)
\tilde{\rho}^{(c)}_{ij}(\epsilon^{\prime})
\tilde{\rho}^{(c)}_{ji}(\epsilon^{\prime\prime}) \chi(\epsilon,
\epsilon^{\prime}, \epsilon^{\prime\prime})}
{z  - \epsilon - \epsilon^{\prime} + \epsilon^{\prime\prime}}~, 
\label{lrpt}
\end{equation}

\noindent with $A_{ii}=[\langle n_{i-\sigma} \rangle (1 - \langle
n_{i-\sigma}\rangle)]/[\langle n_{i-\sigma} \rangle_c (1 - \langle
n_{i-\sigma}\rangle_c)]$ and $A_{i \neq j}=1$. Here $\langle n_{i \sigma}
\rangle_c=\int d\epsilon \tilde{\rho}^{(c)}_{ii}(\epsilon) f(\epsilon)$ with
$f(\epsilon)$ denoting Fermi's distribution. The matrix
$\tilde{\rho}^{(c)}_{ij}(\epsilon)$ describes the density of states of a system
with an empty site $i$ (or sites $i$ and $j$) embedded in a medium with a
coherent potential $\tilde{\Sigma}(z)$. This coherent potential is determined
self-consistently from 
$\tilde{\Sigma}(z)=N^{-1}\sum_{\mbox{\boldmath$k$}} 
\Lambda_{\mbox{\boldmath$k$}}(z)$
where $N$ is the number of sites. Moreover, $\chi(\epsilon, \epsilon',
\epsilon'')=f(-\epsilon) f(-\epsilon') f(\epsilon'') + f(\epsilon) f(\epsilon')
f(-\epsilon'')$. We want to emphasize that in
Eqs. (\ref{lk})-(\ref{icrij}) all memory matrices with site $i$
separated sufficiently far from site $j$ are taken into account until
convergency is obtained.  

It follows from Eqs. (\ref{memcij}) and (\ref{lrpt}) that the self-energy
reduces in the limit of small $U$ to second order perturbation theory while in
the limit of large $U$ the exact result of the atomic limit is reproduced
\cite{kake04-1}. Although the above computational scheme looks at first sight
somewhat difficult to handle, this is not really the case. In fact, it allows
us to calculate the self-energy directly without having to do a numerical
analytic continuation or an interpolation in $\mbox{\boldmath$k$}$ space. Therefore we
obtain for it a high resolution in energy and momentum. In the numerical
calculations we have done, we assumed a paramagnetic ground state since the 
antiferromagnetism is suppressed away from half-filling. Whenever possible we
have made comparisons with Quantum Monte-Carlo (QMC) results and the agreement
was always very satisfactory. 

In Fig. 1 the momentum dependent excitation spectrum is shown in the underdoped
case for $U=8$ (in units of the nearest-neighbor transfer integral) and
$T=0$. One notices an empty upper Hubbard band centered around the $M$ point
and a flat quasiparticle band crossing the Fermi energy $\epsilon_F$. There is
also incoherent spectral density near the $\Gamma$ point resulting from the
lower Hubbard band. Also shown are QMC results for finite temperatures
\cite{grober00}. Results for the Fermi surface are shown in Fig. 2. For
$n=0.95$ (underdoped case) a hole-like Fermi surface is obtained. Due to a
collapse of the lower Hubbard band the portion of the flat band around the $X$
points sinks below the Fermi level. In the overdoped
regime the Fermi surface is electron-like. It is seen that Luttinger's theorem
\cite{luttinger60} does not apply here. These results agree with
the ones obtained from the DCA \cite{maier02}. 

By taking numerical derivatives of $\Lambda_{\mbox{\boldmath$k$}}(z)$ 
we have determined
the momentum dependent effective mass 
$m_{\mbox{\boldmath$k$}}=1-\partial {\rm Re}
\Lambda_{\mbox{\boldmath$k$}}(0^+)/\partial \omega$ 
in the underdoped regime (see
Fig. 3). To our knowledge this is something which could not be done before. For
doping less than 2 \% $m_{\mbox{\boldmath$k$}}$ changes strongly between the minimum value
at the $M$ point and the maximum value at the $X$ point, while
for dopings larger than 2 \% the momentum dependence of 
$m_{\mbox{\boldmath$k$}}$ is weak
with a maximum at $\Gamma$ and not at $X$ as in the underdoped regime. Most
important is the strong dependence of $m_{\mbox{\boldmath$k$}}$ 
near the $X$ point on the
chosen step size $\delta \omega$ when the derivative is taken. This is a clear
signature of MFL behavior. Because of the numerical derivative taken of
Re$\Lambda_{\mbox{\boldmath$k$}}$ at $\omega=0$ we obtained a finite
value of 
$m_{\mbox{\boldmath$k$}}\sim {\rm ln} \delta \omega$ 
instead of a divergency. In fact, in the limit
of vanishing hole doping (half-filled case) we find a typical MFL behavior of
$\Lambda_{\mbox{\boldmath$k$}}(z)$ 
for $U=8$ and $\mbox{\boldmath$k$}=(\pi/2, \pi/2)$ like in the weak
interaction limit. This is shown in Fig. 4 where the two cases are
compared. Because of these features we conclude that for $U=8$ and doping less
than 2 \% MFL theory applies, while for doping concentrations of more than 2 \%
normal Fermi liquid theory is valid.

The different nature of the states for doping concentrations $\delta_h < 0.02$
and $\delta_h > 0.02$ is clearly seen in the density of states (DOS) presented
in Fig. 5 for $U=8$. Consider first the case $\delta_h < 0.02$. With increasing
doping concentration spectral density is shifted from the lower to the upper
Hubbard band, or more generally to higher energies. As a consequence the peak
in the DOS remains at $\epsilon_F$, i.e., it does not shift for small doping
concentrations. Therefore the self-energy has to good approximations the same
frequency dependence as for half-filling. This is the origin of MFL behavior.
When $\delta_h > 0.02$  the lower Hubbard band has essentially collapsed and
the peak in the DOS moves away from $\epsilon_F$. In that case conventional
Fermi liquid behavior sets in. This occurs in a rather discontinuous way. This
is seen from the phase diagram shown in Fig. 6. The MFL regime with a still
existing lower Hubbard band is separated by a region in which two
self-consistent solutions are found from the one with a collapsed lower Hubbard
band. One may consider that transition as one from quasi-localized electrons to
fully itinerant ones. Note that when $\delta_h < 0.01$ the MFL regime is
divided into two regimes. The two solutions found are only slightly different
in the amount of the reduction of the lower Hubbard band. For fixed value of
$\delta_h$ the discontinuous behavior of the self-energy as function of $U$
implies that Luttinger's theorem \cite{luttinger60} is not applicable. 
For $U<6.5$ the MFL behavior at half-filling
changes smoothly to a Fermi liquid state. 

In summary, by using the SCPM we could calculate the zero-temperature 
self-energy $\Lambda_{\mbox{\boldmath$k$}}(z)$ 
for the 2D Hubbard model with high
resolution with respect to $\omega$ and $\mbox{\boldmath$k$}$. 
We find a MFL like behavior
for quite a large range of hole doping concentration and $U$. We obtain there a
strong momentum dependence of the effective mass. In cases where a
comparison with finite temperature results
\cite{bulut94,preuss94,grober00,maier02} can be made the agreement is
good. When $U>6.5$ a discontinuous change takes place with increasing hole
concentration from more localized electrons with a lower Hubbard band to fully
itinerant ones. In the latter case the lower Hubbard band is absent. It is
precisely the transfer of spectral density from the lower Hubbard band to
higher energies which results in MFL behavior at low hole concentrations. Above
the upper discontinuity lines Luttinger's theorem is not applicable. Very close
to half-filling long-ranged antiferromagnetic correlations are expected to
modify the present results. Those correlations have been neglected here. 



\clearpage

\begin{figure}
\caption{Single-particle excitation spectra along high symmetry line for $U=8$
and $T=0$ in unit of the nearest-neighbor transfer integral: electron occupation number $n=0.95$.  Open circles with error bars are the QMC results 
\cite{grober00} at $T=0.33$. The dashed curves show the Hartree-Fock
contribution $\epsilon_{k}$.}
\label{figspec}
\end{figure}
\begin{figure}
\caption{Excitation spectra at the Fermi energy showing Fermi surfaces for
$n$=0.95 and 0.80.  A 80$\times$80 mesh was used for calculations in the
 Brillouin zone} 
\label{figfs}
\end{figure}
\begin{figure}
\caption{Momentum-dependent effective mass $m_{\bf k}$ as a function of doping
concentration. Closed circles: maximum value of $m_{\bf k}$ at $X (\pi,0)$ for
the hole concentration $\delta_{h}=1-n \le 0.02$ and 
$\Gamma \ $(0,0) for $0.02 \le \delta_{h}$, open circles: average
$m_{\bf k}$, closed triangles: minimum value at $M (\pi,\pi)$. Numerical
derivatives are taken with respect to energy fraction
 $\delta\omega=0.05$.  For the
maximum $m_{\bf k}$, also results for $\delta\omega=0.005$ are shown (+). The
momentum-independent effective mass in the single-site approximation (SSA) is
shown by the dashed curve.} 
\label{figmass}
\end{figure}
\begin{figure}
\caption{Real part (thin solid line) and imaginary part (solid line) of the
self-energy at the $k$ point $(\pi/2,\pi/2)$ and half filling. Corresponding
results of second-order perturbation theory are shown by dashed lines.}
\label{figself}
\end{figure}
\begin{figure}
\caption{Densities of states (DOS) for the two self-consistent solutions 
at $n=0.98$.  One $n=0.98+$ (solid curve) is smoothly connected 
with the region $n \ge 0.98$, while the other $n=0.98-$ (dot-dashed
 curve) with the region $n \le 0.98$.  The DOS for $n=1$ is also shown (dotted
line). Note that $\delta=0.02$ (imaginary part of energy) was used in the numerical calculations, so that the peak at $\omega=0$ remains finite for $n=1$.}
\label{figdos}
\end{figure}
\begin{figure}
\caption{Phase diagram showing discontinuity lines. The different transition
lines are explained in the text. The regimes with two self-consistent
solutions are shown by hatched areas. Dashed lines are extrapolations.} 
\label{figline}
\end{figure}


\begin{thebibliography}{1}
\bibitem{Lambda}
We adoped here the self-energy with parameters $(\lambda,\lambda')$= (0,1) in
Eq. (53) of Ref. \cite{kake04-1}.
\end{thebibliography}
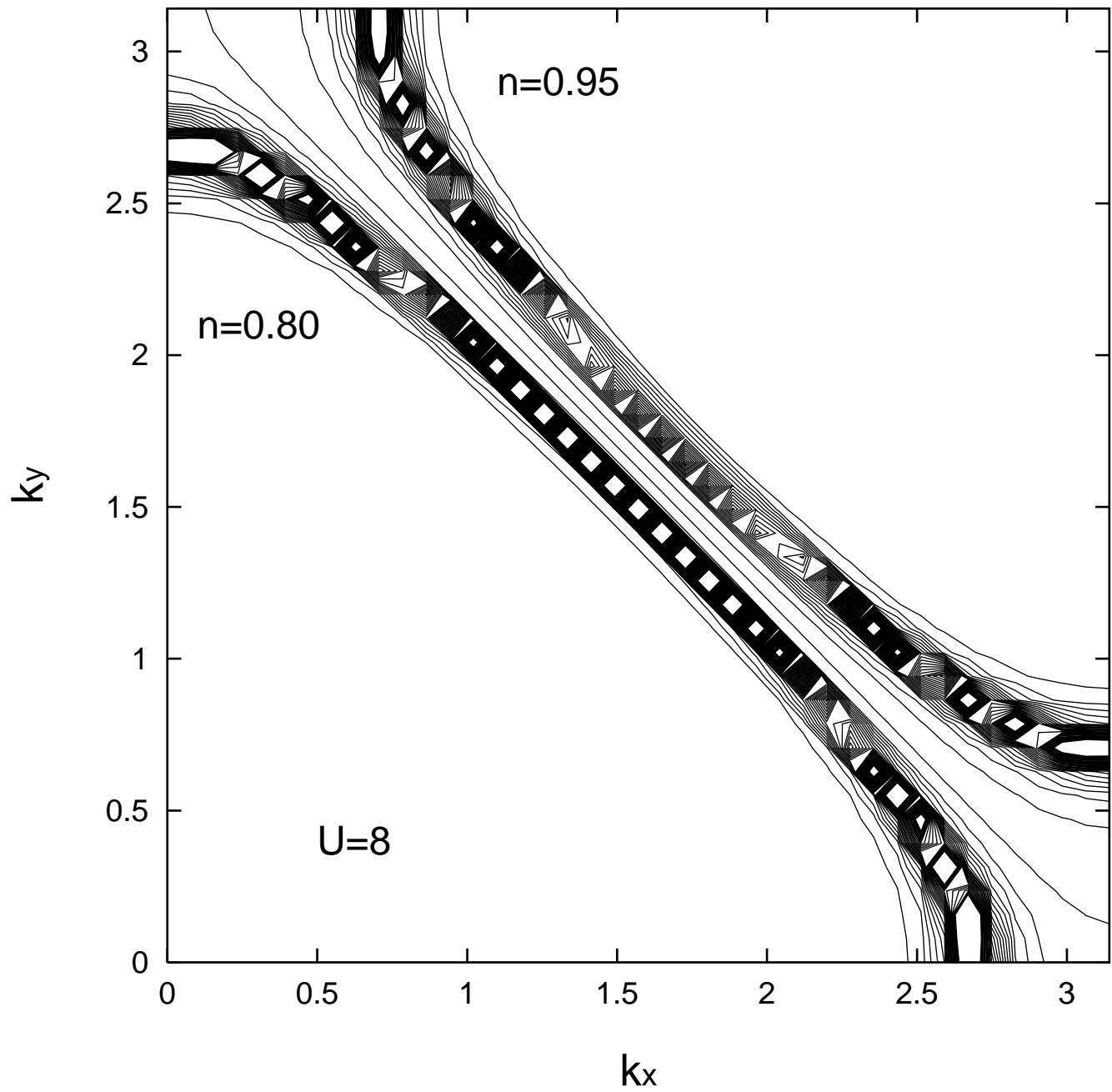
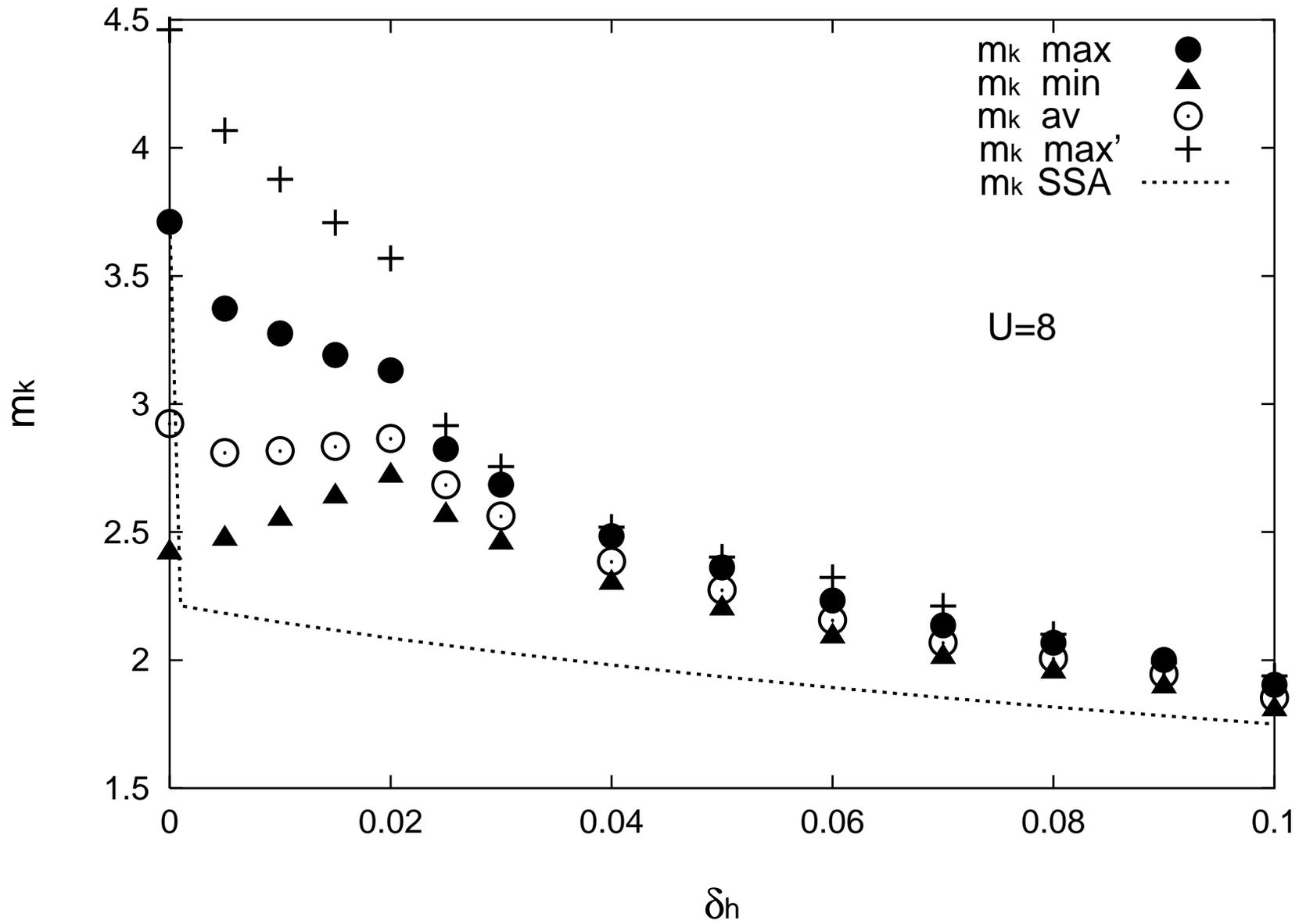
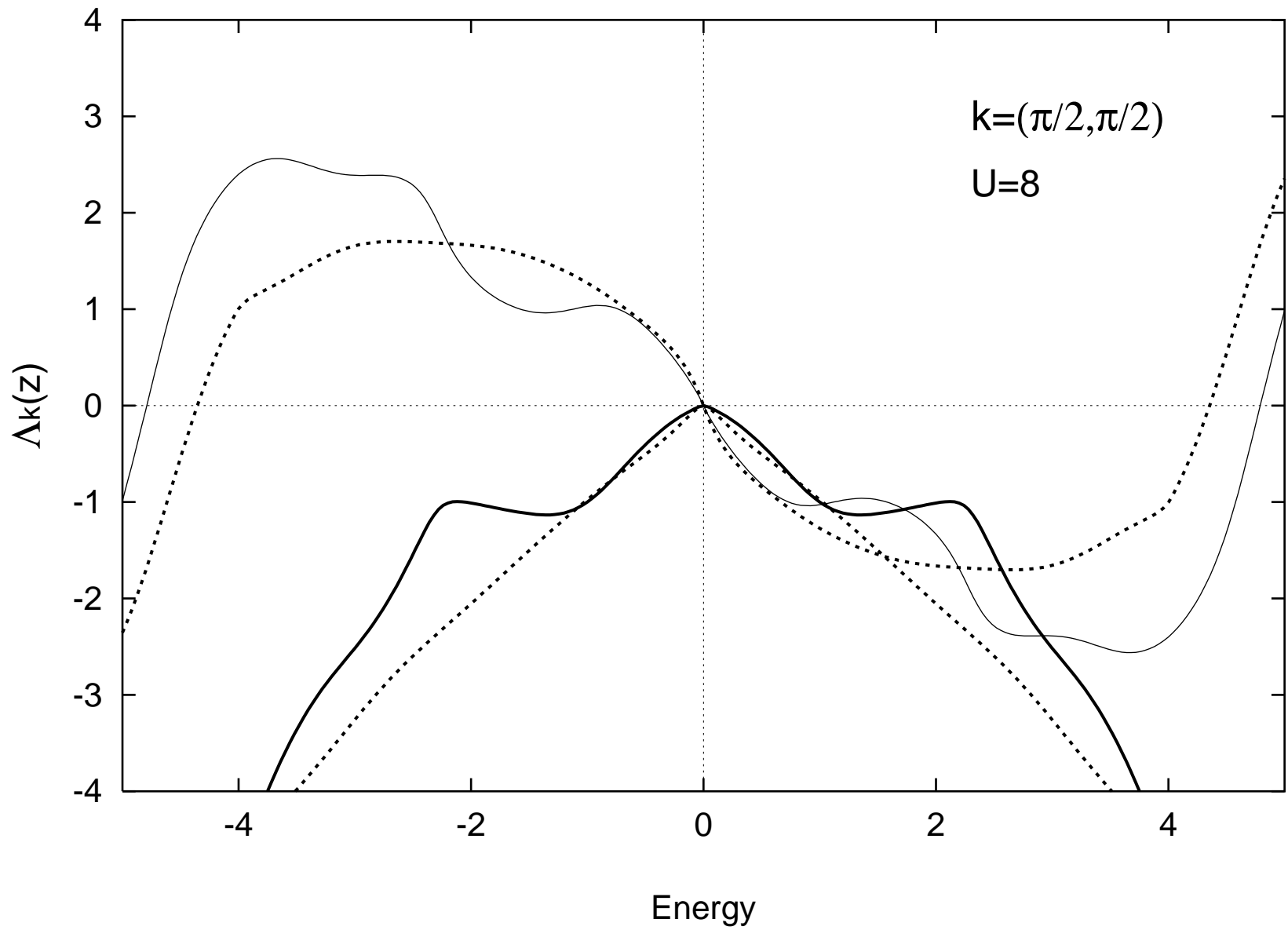
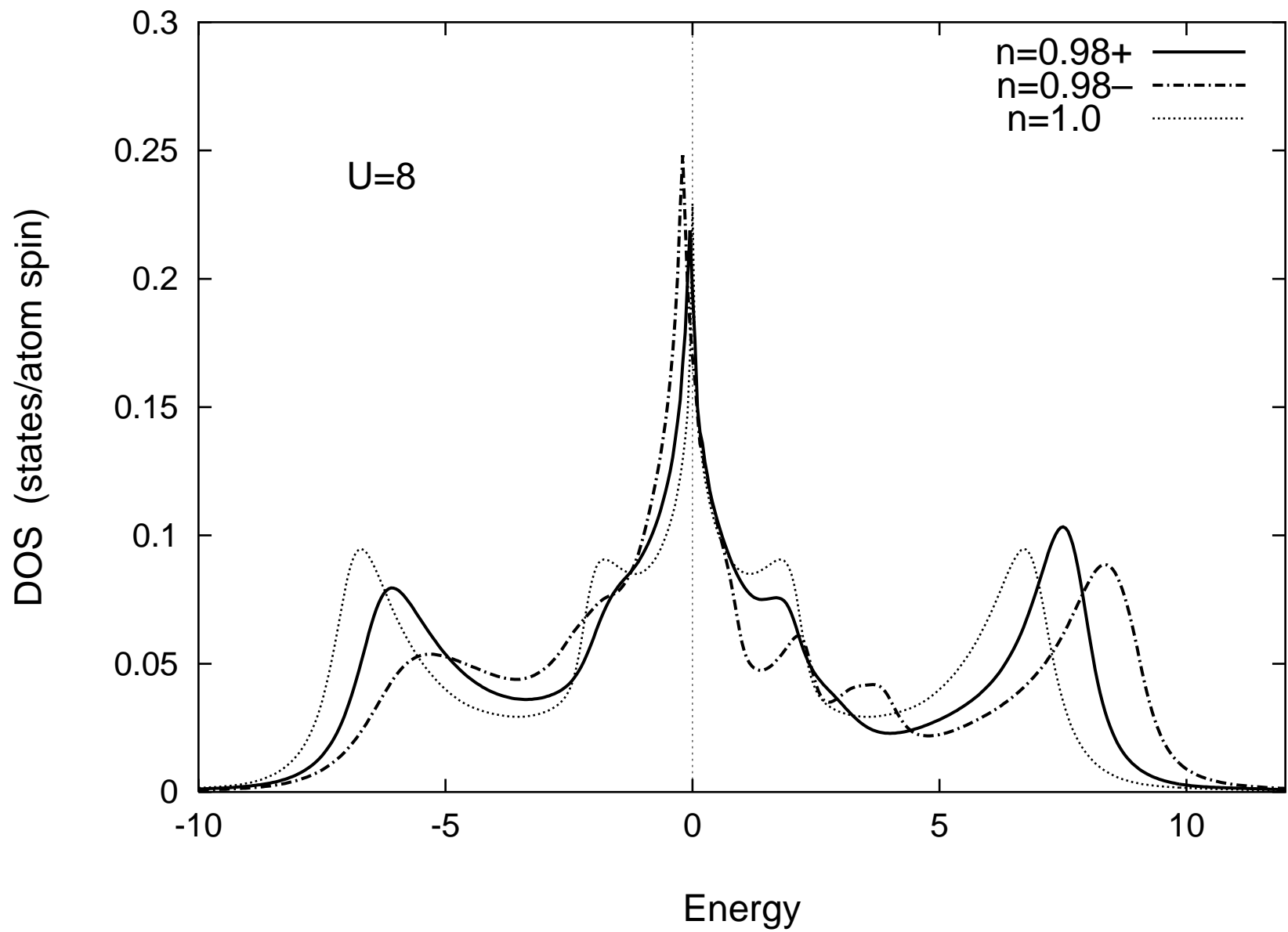
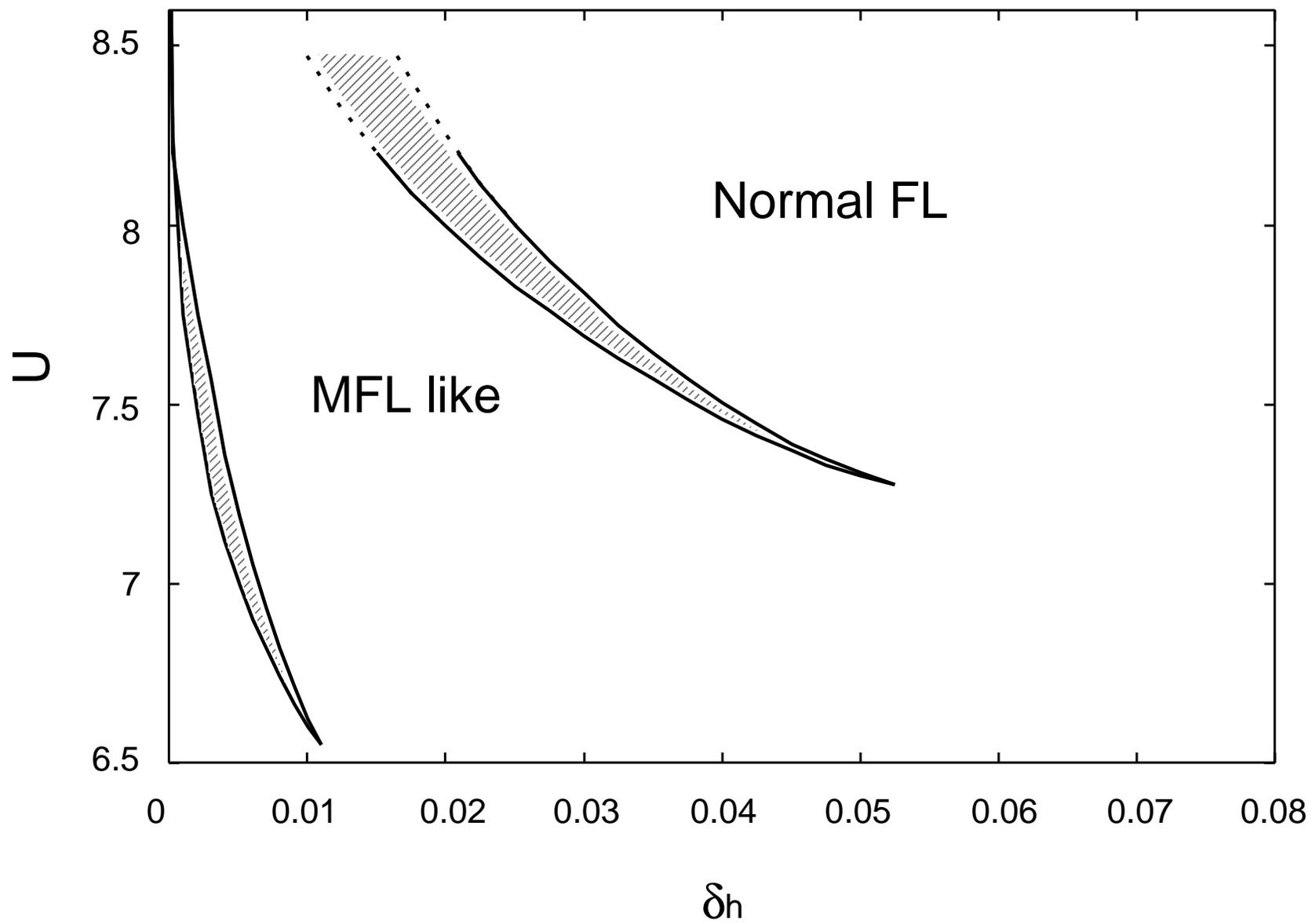
\end{document}